\documentclass[superscriptaddress,journal=aps,twocolumn,babel]{revtex4-1}
\usepackage{amsmath}
\usepackage{amssymb}
\usepackage{siunitx}
\usepackage{amsthm}
\usepackage{physics}
\usepackage{verbatim}
\usepackage{graphicx}
\usepackage{appendix}

\usepackage{times}	
\usepackage{tikz}
\usepackage{soul}

\begin{document}

\title{Reducing qubit requirements for quantum simulation using molecular point group symmetries}

 \author{Kanav Setia}
 \email{kanav.setia.gr@dartmouth.edu}
 \affiliation{Department of Physics and Astronomy, Dartmouth College, Hanover, NH 03755, USA}
  \affiliation{IBM T.J. Watson Research Center, Yorktown Heights, NY 10598, USA}

\author{Richard Chen}
\affiliation{IBM T.J. Watson Research Center, Yorktown Heights, NY 10598, USA}
 
\author{Julia E. Rice}
\affiliation{IBM Almaden Research Center, 650 Harry Road, San Jose, CA 95120, USA}
 
\author{Antonio Mezzacapo}
\affiliation{IBM T.J. Watson Research Center, Yorktown Heights, NY 10598, USA}
 
\author{Marco Pistoia}
\affiliation{JP Morgan Chase \& Co.}
 
\author{James D. Whitfield}
 \affiliation{Department of Physics and Astronomy, Dartmouth College, Hanover, NH 03755, USA}
 
\date{\today}

\begin{abstract}
Simulating molecules is believed to be one of the early-stage applications for quantum computers. Current state-of-the-art quantum computers are limited in size and coherence, therefore optimizing resources to execute quantum algorithms is crucial. In this work, we develop the second quantization representation of the spatial-symmetries which are then transformed to their qubit operator representation. These qubit operator representations are used to reduce the number of qubits required for simulating molecules.  We present our results for various molecules and elucidate a formal connection of this work with a previous technique that analyzed generic $Z_2$ Pauli symmetries.
\end{abstract}
\maketitle
\section{Introduction}
Quantum simulation of chemistry is one of the most promising applications for near-term quantum computers. The interest in the field has grown exceptionally,  
resulting in many
algorithms for quantum simulation on quantum computers. Of particular interest have been the improvements in algorithms for near-term noisy intermediate-scale quantum (NISQ) devices \cite{Preskill2018}. 

There has been constant improvement in the resources required for quantum simulation. Prohibitively large number of gates required for the evolution of wave functions \cite{Wecker2014} in the case of phase estimation algorithm led to the development of hybrid algorithms like variational quantum eigensolver \cite{Peruzzo2014}.

At the same time, multiple error mitigation techniques for NISQ devices have been proposed \cite{Bonet2018,Temme2017}. Some of these techniques involve extrapolating errors \cite{Kandala2019} and make use of the symmetries present in the Hamiltonian \cite{Bonet2018,Mcardle2019} to partially correct errors. There have been some recent work that involved using different fermionic encodings \cite{Setia2018} to reduce the number of gates and get some error mitigation. The number of qubits required for such encodings are greater than the number of fermionic modes in the system.  We refer readers to references \cite{Mcardle2018, Cao2018} for a review of the various techniques for quantum chemistry on quantum computers.

These developments have contributed towards the simulation of molecules such as beryllium hydride on a quantum computer \cite{Kandala2017}. Further, there has been a lot of development in the quantum computer architecture other than the superconducting qubits, e.g. trapped ion quantum devices \cite{Nam2019}. 

Even with all these exciting developments, we are still some time away from the fault tolerant quantum computers. The qubits remain precious resources for the NISQ devices and it is important to keep minimizing the number of qubits required for simulating a particular system. In this work, we present techniques to use the symmetries present in the molecules to reduce the number of qubits required for the simulation.

In ref. \cite{Bravyi2017}, a procedure for tapering off qubits based on $Z_2$ symmetries was developed. The idea involved finding a Pauli string that commutes with the Hamiltonian. An efficient algorithm was presented to find Pauli strings which commute with the Hamiltonian. Such Pauli strings/operators are called the symmetries of the Hamiltonian. Based on these Pauli strings, a unitary operator could be found which transforms the Hamiltonian in such a way that the Hamiltonian acts trivially or at most with $\sigma_x$ on a set of qubits. The qubits on which Hamiltonian acts trivially or with $\sigma_x$ can then be left out of the simulation. Effectively, the Hamiltonian has been projected into symmetry sub-spaces all of which can be simulated using lesser number of qubits. 
Based on the results in previous paper~\cite{Bravyi2017}, it is natural to ask the physical meaning of the obtained symmetries. For few of the Pauli strings (symmetries), it is easy to figure out the physical meaning, e.g. if Jordan-Wigner transform was used then the string $Z^{\otimes N}$ corresponds to the parity of the fermions. For many other Pauli strings, the correspondence to a physical symmetry was not obvious. 

In our current work, we have developed a collection of new techniques. These enable us to write down the second quantization representation of the point group symmetries. This second quantization representation could then be transformed to qubit operator representation. In general, this qubit operator representation is a summation of the Pauli strings. We then introduce a technique to turn the summation of the Pauli-strings representation of a subset of these symmetries to a single Pauli-string. Some of the symmetries obtained are the same as the ones found in ref. \cite{Bravyi2017}. Therefore, our work also provides the physical meaning for the symmetries found in the previous work. The commuting set of symmetries which are represented by single Pauli-strings can be used to reduce the qubit count using the tapering off qubit procedure (Section \ref{Sec:tap_of_q}). We also present a way to pick out the correct eigensector of the Hamiltonian. Knowing the correct eigensector beforehand reduces the number of experiments to be run compared to the previous work. This improvement is exponential in the number of qubits tapered e.g.  if five qubits are tapered, then instead of running the experiment 32 times (because we do not know which eigensector contains the ground state), we just need to run it once \cite{Setia2020}. 

An outline of the paper is as follows:
in Section \ref{Sec:tap_of_q}, we review a result on qubit tapering of second-quantized Hamiltonians presented in~\cite{Bravyi2017}. Building on that, in Section \ref{Sec:PGS} and \ref{Sec:PG_SQ} we discuss point group symmetries and formalize a second quantization representation of point group symmetries in molecular systems. In Section \ref{Sec:Trans_PG_to_SP}, we present the procedure to diagonalize the matrix representation of the second quantization representation. These diagonal matrices transform to a single Pauli-string. These Pauli operators can then be used to taper off qubits using results from~\cite{Bravyi2017}. Finally, in Section \ref{results}, we present our results for different molecules belonging to different point groups. 
\break
\break

\section{Qubit tapering based on ${Z}_2$ symmetries} \label{Sec:tap_of_q}
A system with $M$ fermionic modes can be described by the following Hamiltonian:
\begin{align}
    \mathcal{H}=\sum_{ij}^{M^2} h_{ij}a_i^{\dagger}a_j+\frac12 \sum_{ijkl}^{M^4} h_{ijkl}a_i^{\dagger}a_j^{\dagger}a_ka_l \label{eq:ham}
\end{align}
where, $h_{ij}$ and $h_{ijkl}$ are the one-body and two-body integrals. For a given molecule, these can be obtained from various quantum chemistry software packages \cite{Psi42017, PySCF2017}. $\lbrace a_i^{\dagger}, a_j^{\dagger}...\rbrace$ and $\lbrace a_i, a_j...\rbrace$ are the creation and annihilation operators which obey the canonical commutation relations:
\begin{align}
    a_ia_j+a_ja_i= 0 \qquad \qquad
    a_ia^{\dagger}_j+a^{\dagger}_ja_i=\delta_{i,j}I. \label{eq:sec_quantization}
\end{align} 
To simulate eq.~\eqref{eq:ham} on a quantum computer we need to map the Hamiltonian to qubit operators. This can be achieved with one of the many transformations available, e.g. the Jordan-Wigner transformation, parity, etc. The transformed fermionic Hamiltonian takes the following form:
\begin{align}
  H=\sum_j^r c_j\boldsymbol{\eta_j} \label{eq:ham_qub}   
\end{align}
where  $r$ is the total number of terms, $\boldsymbol{\eta_j} \in \mathcal{P}_M$ and $\mathcal{P}_M$ is  Pauli group given by:
\begin{align}
    \mathcal{P}_M=\pm \lbrace I,\sigma_x,\sigma_y,\sigma_z\rbrace ^{\otimes M}
\end{align}
 
For the Jordan-Wigner, parity, and Bravyi-Kitaev \cite{Bravyi2000} transformation (encoding), the length of the string is $M$, same as the number of modes, but this could be different for mappings like superfast encoding (SE) and generalized superfast encoding (GSE) \cite{Setia2018}.

Consider a set of $k$-qubits out of total $M$, on which all the terms, $\{\boldsymbol{\eta_j}\}$, in the Hamiltonian act trivially (with identity operator), then it is easy to see that we do not need to include those $k$-qubits in the simulation. Further, the qubits can still be left out even if all the terms, $\boldsymbol{\eta_j}$ act on the $k$-qubits with at most one Pauli-gate, e.g $\sigma_x$. In such a scenario, the single qubit Pauli-gate appearing in various $\boldsymbol{\eta_j}$ terms can be replaced by their eigenvalues, $\pm 1$, so that the $j$-th qubit can be tapered off.

So, we are motivated to transform the Hamiltonian in such a way that it acts trivially or at most with one Pauli-gate on a subset of qubits. For this, we make use of the symmetries present in the system. Consider an abelian group $\mathcal{S}\in \mathcal{P}_M$, such that $-I\not\in \mathcal{S}$. Such a group is called the symmetry of the Hamiltonian if all the elements of $\mathcal{S}$ commute with each Pauli term of the Hamiltonian. We know that every group has a set of generators, $\lbrace \boldsymbol{\tau}_1,..,\boldsymbol{\tau}_k\rbrace $, and from stabilizer theory \cite{Gottesman1997,Neilsen1998} we know:

\begin{align}
U_i\boldsymbol{\tau}_i U_i^{\dagger} = \sigma_x^{q(i)} \qquad q=\{a,b,...\} \label{eq: UtU}
\end{align}
where, $U_i\in \mathcal{C}_M$. Clifford group \cite{Gottesman1998}, $\mathcal{C}_M$, on $M$-qubit is defined as the set of unitary operators, $U$, such that $$U\boldsymbol{\gamma}U^{\dagger}\in \mathcal{P}_M\qquad \forall \boldsymbol{\gamma}\in \mathcal{P}_M$$
If we could find the group of symmetries of the Hamiltonian, then we could transform the generator set of the symmetries to single qubit Pauli operators using eq. \eqref{eq: UtU}. For a given symmetry, if we transform the Hamiltonian using the unitary, $U_i$, in eq. \eqref{eq: UtU}, then each Pauli term in the transformed Hamiltonian, $U_iHU_i^{\dagger}$ must commute with $\sigma_x^{q(i)}$, i.e 

$$U_iHU_i^{\dagger}=\sum_j c_j\boldsymbol{\sigma_j}\qquad[\boldsymbol{\sigma_j},\sigma_x^{q(i)}]=0$$
where, $\boldsymbol{\sigma_j}=U_i\boldsymbol{\eta_j}U_i^{\dagger}$. 
This would imply that the transformed Hamiltonian must be acting trivially or at most with $\sigma_x$ on the $q(i)^{th}$ qubit. This would allow us to replace the $\sigma_x$ on $q(i)^{th}$ qubit by its eigenvalue, and remove the qubits from the simulation.

We now describe two important sub-procedures for the above technique to work. First, we describe how to find the symmetry group and then we will describe the procedure to find the unitary, $U$, to transform the generating set of the symmetries as well as the Hamiltonian.

\emph{Finding symmetries:} A Pauli string, $\boldsymbol{\eta}$, acting on $N$ qubits can be parameterized by a binary string $(a_x|a_z)$, of length, $2N$ where each component of vectors $a_x$ and $a_z$ are zero or one \cite{Dehaene2003}. 
Qiskit uses the same representation for the Pauli class to represent a Pauli string. This way each $\boldsymbol{\eta}(a_x|a_z)$ can be represented as:
\begin{align*}
    \boldsymbol{\eta}(a_x|a_z)=\prod_{i\in a_x}\sigma_x^i \cdot \prod_{j\in a_z}\sigma_z^j
\end{align*}
This parameterization is very effective when we need to multiply two Pauli strings, or if want to check whether the terms commute.
\begin{align*}
    \boldsymbol{\eta}(a_x|a_z)\boldsymbol{\eta}(b_x|b_z) = (-1)^{a_x b_z+a_z b_x}\boldsymbol{\eta}(b_x|b_z)\boldsymbol{\eta}(a_x|a_z).
\end{align*}
In order for the terms to commute, $a_x b_z+a_z b_x=0 \mod 2$. 

We can represent all the Pauli strings appearing in the Hamiltonian by a binary matrix:

\begin{align}
    G(H)=\left[\frac{G_x}{G_z}\right]
\end{align}
where $j^{th}$ column of $G$ is a binary matrix corresponding to ($a_x|a_z$) representing $\boldsymbol{\eta}_j$. It can be seen that the size of the $G$ matrix will be $2M\cross r$, where $r$ is the total number of terms in the Hamiltonian.
From the $G$ matrix, we can construct another check matrix, $E$:
\begin{align}
    E=[(G_z)^T|(G_x)^T]
\end{align}
It can be observed that the kernel of the check matrix, $E$, gives the elements of the symmetry group. Using the $Ker(E)$, one can obtain the generators, $\boldsymbol{\tau}_i$, of the group by using the Gram-Schmidt orthogonalization procedure over the binary field, $Z_2$.

\emph{Finding the unitaries:}
Once we have the generators of the symmetries $\left\lbrace\boldsymbol{\tau}_i\in \mathcal{S} \right\rbrace$ of the Hamiltonian then, as discussed above, each of these symmetries can be turned into a Pauli-$X$ operator on a single qubit using eq. \eqref{eq: UtU}. To find the unitary, $U$, we try to find the $\sigma_x$ on a qubit such that it anti-commutes with one of the symmetries and commutes with all the other symmetries. Then,
$$U_i=\frac{1}{\sqrt{2}}(\tau_i+\sigma_{q(i)}^x)$$
Further, we can use the permutation operators to bring qubits belonging to set $q$, to the end. 
These $U_i$s along with permutation operators $W_i$s can then be used to transform the Hamiltonian. 
$$(U_1W_1U_2W_2...U_kW_k)H(W_k^{\dagger}U_k^{\dagger}...W_1^{\dagger}U_1^{\dagger})=\sum_j c_j\boldsymbol{\sigma_j}$$
where $W_i$s are the permutation matrices. The transformed Hamiltonian now commutes with $\sigma_x^{q(i)}$. This implies that all the terms in the Hamiltonian must act trivially on the last $k$-qubits or with just $\sigma_x$'s. In case of variational quantum eigensolver algorithm, we can therefore get rid of the last $k$ qubits and the  $\sigma_x$ operators are replaced with their eigenvalues $\pm 1$.

This tapering off qubit procedure was shown to find many symmetries of molecular systems \cite{Bravyi2017}. Bravyi et. al. used the procedure on Hamiltonians of many molecules such as $H_2O$, $LiH$, $BeH_2$, etc. For all the molecules, standard geometry and STO-3G basis set were used. Further, many different fermionic encodings, like the Jordan-Wigner, Bravyi-Kitaev, and parity, were explored to transform the fermionic Hamiltonian. The different encodings did not affect the total number of qubits tapered. For $LiH$, $H_2O$, and $BeH_2$, 4, 4, and 3 qubits were tapered off, respectively. 

The question that was left unanswered and is also the inspiration for the present work was `what is the physical significance of the symmetries obtained using tapered off qubit procedure?'. Further, we are also interested in exploring their connection to the point group symmetries of the molecules.

\section{Point group symmetries}\label{Sec:PGS}
For describing point-group symmetries in molecules only four types of non-trivial point-group symmetry operations are required \cite{Cotton1997}. 
These are, 
\begin{itemize}
    \item[1)]\emph{Proper rotation (${C_n}$)} is a rotation by $360/n$ degrees.
    \item[2)]\emph{Plane-reflection (${\sigma}$)} is a reflection in  a given plane.
    \item[3)]\emph{Improper axis rotation (${S_n}$)} is a rotation by $360/n$ degrees followed by reflection in plane perpendicular to the rotation axis.
    \item[4)]\emph{Center of inversion (${i}$)} is the inversion ($\vec x \rightarrow -\vec x$) of all atomic coordinates about the center.
\end{itemize}
It can be shown that center of inversion ($i$) is actually same as $S_2$.

Further, if we consider all the symmetry elements of a given molecule, then it can be proven that they satisfy the axioms of a group. All four symmetries leave at least one point unchanged in space and hence the name point group symmetry.

Molecules belong to different point groups based on the different symmetry elements that leave the molecule unchanged. For example, water ($H_2O$) belongs to $C_{2v}$ group because it has a vertical proper rotation axis ($C_2$) and a plane of reflection in the symmetry group, and ammonia, $NH_3$, belongs to $C_{3v}$ group because it has a vertical proper rotation axis $C_3$ and a vertical plane of reflection. The groups are represented using character tables, e.g. character table for group $C_{3v}$ is given below:
\\
\[
\begin{tabular}{ccrr}
     \phantom{sc}$C_{3v}$\phantom{sc}&\phantom{spc}E\phantom{spc} & \phantom{spc}$2C_3$ & \phantom{spc}$3\sigma_v$\\ 
     \hline
     \hline
     $A_1$& 1 & 1& 1    \\
     $A_2$& 1 & 1& -1 \\   
     E & 2 & -1 & 0 \\        
\end{tabular}
\]
The entries of the table are the characters of symmetry operations  within different irreducible representations.  For each irreducible representation, the trace of the matrix representation is called the character.  The labels for each row correspond to different irreducible representation and the symmetry elements are grouped into classes matching each column \cite{Cotton1997}.

There are many other valid representations of the group. One of the ways to build a representation is to consider a vector representation of a point in space and build matrices that transform the vector according to the group action. Another way, which is used to build symmetry adapted linear combinations (SLACs), is to consider a vector representation of atomic orbital functions or molecular orbital functions and build matrices that transform the vector according to the group action. Traditionally, such a procedure is used to build symmetry adapted molecular orbitals from the atomic orbitals. We show that such a representation could also be used to taper off qubits.

\section{Point group symmetries in second quantization}\label{Sec:PG_SQ}

Our aim is to explore the relation of point group symmetries to the symmetries found using tapering off qubit procedure. For this, we first develop the second quantization representation of the point group symmetries. We start by considering a finite set of the single particle wave functions given by
\begin{align}
\left\lbrace\phi_i(x), i\in [1,M]\right\rbrace.   
\end{align}

These functions form a basis set which becomes complete as $M\xrightarrow{}\infty$. As it is computationally very expensive to deal with large basis sets, truncated basis sets are used. 
Now, assume the system under consideration has some point group symmetry. If $\boldsymbol{R}$ is the operator that defines the symmetry operation then we get:
\begin{equation*}
    \phi'_i(x)=\boldsymbol{R}(\phi_i(x))
\end{equation*}

It is possible to pick the truncated basis set in such a way that $\boldsymbol{R}$ ends up being a linear transformation, $R$, which gives us:
$$\phi'_i(x)=\sum R_{ji} \phi_j(x)$$
Further, we require that $R$ be a unitary matrix, so that the transformed second quantization operators still satisfy the canonical commutation relations \eqref{eq:sec_quantization}. In second quantization formalism, the $R$ matrix is given by:
\begin{align}
    b_i=\sum_{j=1}^M R_{ji} a_j
\end{align}
where, $b_i$ and $a_j$ are the second quantization operators associated with $\phi'_i(x)$ and $\phi_j(x)$, respectively.
For a given molecule, we can find an $M\cross M$ matrix representation for each symmetry operation, $R$. The matrices will then follow the multiplication table of the symmetry group.

The one-body integrals, $\{h_{ij}\}$, 
can be represented by an $M\cross M$ matrix and the two-body integrals, $\{h_{ijkl}\}$, 
can be represented by $M\cross M\cross M \cross M$ tensor. We can use the unitary matrix $R$ to transform the one-body and two-body tensors and check whether the Hamiltonian remains the same. This is same as checking the commutator of Hamiltonian with the symmetry. The commutation of $R$ matrix with the Hamiltonian verifies that it is a symmetry.

It is important to note that we are able to check the commutation of $R$ matrix with the Hamiltonian without going to the $2^M \cross 2^M$ representation of the Hamiltonian. Given the matrix for the unitary $R$, it is not trivial to get its second quantization representation. But, if we restrict the unitary to be signed permutation matrices then the second quantization representation could be constructed easily. For example, the second quantization representation for swapping, say, mode-p and mode-q is given by:

\begin{align}
    \mathcal{R}_{pq}=I-a_p^{\dagger}a_p-a_q^{\dagger}a_q+a_p^{\dagger}a_q+a_q^{\dagger}a_p
\end{align}
It can be shown that any permutation can be decomposed into a series of transpositions each swapping two elements at a time. The second quantization operator that gets us the (-1) phase is given by $1-2a_p^{\dagger}a_p$. This operator, along with the second quantization operator for the permutation matrix, lets us generate second quantization representation of any signed permutation matrix.

It can be seen that the $R$ matrices can be constructed using just the signed permutation matrices if the basis sets chosen are atom-centred basis sets. This, in fact is a common choice while performing molecular calculations.

With the second quantization representation available for the symmetries, we can use any fermion-qubit transformation to get a qubit operator representation of $R$ matrix. In general, the qubit representation of $R$ ends up being a summation of Pauli strings. The procedure presented in Section \ref{Sec:tap_of_q} can be used if the symmetry is a single Pauli string. In the following section, we present the conditions under which this could be achieved as well as the procedure to do so.

\section{Representing point group symmetries as Pauli operators} \label{Sec:Trans_PG_to_SP}

Consider the Hamiltonian in eq. \eqref{eq:ham}. Following the section \ref{Sec:PG_SQ}, we consider the symmetries of the Hamiltonian represented by signed permutation matrix. Suppose $\boldsymbol{\pi}$ is a permutation under which Hamiltonian is invariant. We show below that in such a case we can get rid of a qubit. 

Suppose $R$ is a unitary matrix of size $n\cross n$. Then there exists an $n$-qubit unitary matrix $\hat{R}$ such that
\begin{align}
    \hat{R}a_p\hat{R}^{\dagger}=\sum_{q=1}^n\bra{p}R\ket{q}a_q \label{eq:Fact1}
\end{align}
for all $1\leq p\leq n$.

Define a permutation matrix $R$ of size $n\times n$ such that 
\begin{align}
    R\ket{p}=\ket{\boldsymbol{\pi}(p)}
\end{align}
for all $1 \leq p \leq n$. 
By assumption, 
\begin{equation}
    \hat{R}H\hat{R}^{\dagger}=H
\end{equation}

Now, since $R$ is unitary, it can be written as
$$R = \exp(iG)$$  for some hermitian matrix $G$ of size $n\cross n$. 
The $n$-qubit unitaries for $R$ and $G$ are given by: 
\begin{align}
\hat{R} = \exp(i \hat{G}),\qquad \hat{G} = \sum_{p,q=1}^n \bra{p}G\ket{q} a^{\dagger}_p a_q
\end{align}

Choose a unitary $n\cross n$ matrix $V$ that diagonalizes $G$ such that 
\begin{equation}
V^{\dagger}GV =\sum_{p=1}^{n}\lambda_p\ket{p}\bra{p}    
\end{equation}
for some real eigenvalues $\lambda_p$.
Let $\hat{V}$ be the $n$-qubit unitary matrix constructed from $V$ following eq.~(\ref{eq:Fact1}). Next, define a new symmetry operator
\begin{equation}
S \equiv \hat{V}\hat{R}\hat{V}^{\dagger}= \exp( \hat{V}\hat{G}\hat{V}^{\dagger})=\prod_{p=1}^n\exp(i\lambda_p a_p^{\dagger}a_p)
\end{equation}
 Note that $
 SH' = H'S, \text{ where } H' = \hat{V}H\hat{V}^{\dagger}     
 $.
 In other words, $S$ is a symmetry of $H'$. Let us now assume that $\boldsymbol{\pi}$ swaps some pairs of modes. Then $\boldsymbol{\pi}^2$ is the identity permutation. Thus $R^2 = I$ which is possible only if $G$ has eigenvalues $\lambda_p\in\left\{0,\pi\right\}$. Let $M$ be the subset of modes $p$ such that $\lambda_p = \pi$. Then 
 $$S =\prod_{p\in M} ( -1)^{a^{\dagger}_pa_p}$$
 
 If we use the Jordan-Wigner encoding of fermions into qubits then $S$ becomes a $Z$-type Pauli operator. Thus we can simulate $H'$ using a system of $n-1$ qubits by exploiting the Pauli symmetry $S$. Finally, if $H$ includes only single-particle and two-particle operators then so does $H'$.
 
Further, it can be realized that we do not need to construct the $G$ matrix, and in fact, we can diagonalize the $R$ matrix directly to obtain the $S$ matrix. Similar results were obtained in~\cite{yen2019exact}, where the qubit representation of symmetries was used to build projectors on target symmetry sectors.  
The qubit operator representing $S$ is an operator that acts with $\sigma_z$ on qubits, $j$, where $S(j,j)=-1$.

Another important thing to note is that in case of multiple symmetries, we will be required to simultaneously diagonalize them. Simultaneous diagonalization will only be possible when the symmetries commute. In case of non-commuting symmetries, maximal set of commuting symmetries will be used to taper off qubits. This will become more clear in our discussion of $NH_3$ molecule.

For example, if for a five dimensional $S$ matrix the $-1$ eigenvalues are in position $(2,2)$ and $(4,4)$,  then the Pauli-$Z$ symmetry will be $Z_2Z_4$. 

The techniques presented in Sections \ref{Sec:tap_of_q}, \ref{Sec:PG_SQ} and \ref{Sec:Trans_PG_to_SP}, can be used to develop the following procedure for tapering off qubits.

\emph{Summary of algorithm.}
\begin{itemize}
    \item[1] For a given molecule with a fixed geometry get the Hamiltonian in the atomic orbital basis.
    \item[2] For a given geometry of a molecule the point group symmetries can be found using one of many algorithms available \cite{Cole2001}.
    \item[3] Find the matrix form of the point group symmetries.
    \item[4] Choose the largest abelian group of the symmetries.
    \item[5] Find the single Pauli-string representation of symmetries using techniques in section \ref{Sec:PG_SQ} and \ref{Sec:Trans_PG_to_SP}.
    \item[6] Taper off qubit using technique in section \ref{Sec:tap_of_q}.
\end{itemize} 

As we have mentioned previously, the general qubit operator representation of the symmetries is not a single Pauli-string. This implies that all the spatial symmetries must not have been found using the method of~\cite{Bravyi2017}. We show that this indeed is the case by finding extra symmetries in the molecules.

\section{Results} \label{results} 
Table \ref{tab:qub_red} presents our results for various molecules that we studied along with the number of qubits we were able to taper off. We have open-sourced our code and have made it available on github \cite{Setia2020}. We now discuss three molecules from the table to illustrate three important points. We start with discussing the $H_2$ molecule to illustrate the procedure, then we discuss the case of $BeH_2$ to demonstrate that the spatial symmetries could be used to reduce more qubits than what was possible with~\cite{Bravyi2017}. Third, $NH_3$ molecule is discussed to show that only abelian subgroup of the symmetry point group could be used to taper off qubits.

\subsubsection*{$H_2$ (Symmetry group: $D_{\infty h}$)}
As per the formalism presented in Section \ref{Sec:PG_SQ}, we want to pick the basis set in such a way that the operation corresponding to the spatial symmetries ends up being a permutation matrix.

We consider the hydrogen molecule at bond length of $0.7414$ \AA.
and choose the basis set to be STO-3G, where a single $1s$ orbital is placed on each of the hydrogen atoms. So, both $C_2$ (rotation about z-axis by $180^{\circ}$) and $\sigma(yz)$ (reflection through yz plane) will have the net effect of swapping the two hydrogen atoms. This corresponds to swapping the 1s orbitals and hence the rotation matrix, $R$ is a permutation matrix:
\begin{align*}
\begin{bmatrix}
0 & 1& 0&0\\
1 & 0 &0&0\\
0&0&0&1\\
0&0&1&0
\end{bmatrix}
\end{align*}

It can be checked that the Hamiltonian does remain the same under this permutation of the fermionic modes. We can then diagonalize the $G$ matrix corresponding to the $R$ matrix and taper off the qubit.

\subsubsection*{$BeH_2$ (Symmetry group:$D_{\infty h}$)  }
For beryllium hydride, the total number of qubits that can be tapered off using point group symmetries are five. In contrast, the symmetries that one can find using results from~\cite{Bravyi2017} (Section \ref{Sec:tap_of_q}) are only four. The geometry used is linear with the bond length between beryllium and hydrogen is $1.291$ \AA  ~for both bonds and STO-3G basis set is used.

Given this geometry, the symmetries that could be represented using a generalized permutation matrix are, $\sigma(xy)$ (reflection in xy-plane), $\sigma(yz)$ (reflection in yz-plane), $\sigma(xz)$ (reflection in xz-plane). The qubit operators for the $R$ operator corresponding to $\sigma(xy)$ and $\sigma(xz)$ turn out to be single qubit operators. The final qubit operators for these symmetries match the symmetries found using~\cite{Bravyi2017}. The unitary, $R$, for the symmetry, $\sigma(yz)$ is not diagonal and the qubit operator representation is a sum of Pauli strings. For this reason, the symmetry is not observed using directly results from~\cite{Bravyi2017}. But, the spatial symmetry can be used to taper off qubit using the procedure given in Section \ref{Sec:Trans_PG_to_SP}.

\subsubsection*{$NH_3$, (Symmetry group: $C_{3v}$)}
Ammonia belongs to the symmetry group, $C_{3v}$. In the symmetry group there are two rotation operators and three reflection operators. The two rotation operators form a class and so do the three reflection operators. If the symmetry group were abelian, we could reduced the qubit count by one for each class. Since the symmetry operators in two classes do not commute, we can reduce the qubit count by just one corresponding to one of the symmetries. Further, as per the formalism we developed, the unitary operator $R$, must square to identity. This implies the only choice we have is the reflection operator. It should be noted that this symmetry did not appear using directly the method presented in~\cite{Bravyi2017}.

\begin{table}[]
    \centering
    \begin{tabular}{c c c c c}
         Symmetry & Molecule & Qubits  & Qubits  & Qubits tapered \\
         & &required & tapered& using ref. \cite{Bravyi2017}\\
        $C_{2v}$        & $H_2 O$       &   14  &   4  & 3\\
        $C_{3v}$        & $N H_3$       &   16  &   3  & 2\\
        $D_{2h}$        & $C_2 H_4$     &   28  &   5  &  3\\
        $D_{3h}$        & $BF_3$         &   40  &   5   & 3\\
        $C_{\infty v}$  & $LiH$         &   12  &   4  & 4\\
        \hline           
                        & $CO_2$        &   30  &   5  & 4\\
        $D_{\infty}h$   & $C_2 H_2$      &   24  &   5 & 4\\
                        & $Be H_2$       &   14  &   5 & 4  
    \end{tabular}
    \caption{The above table shows the molecules we tested with our technique and the number of qubits we were able to taper off. This includes the two qubit reduction due to conservation of spin up electrons and the spin down electrons. All the symmetries here form an abelian group. For all the molecules equilibrium geometry and STO-3G basis set is used. The one-body and two-body terms are obtained in atomic orbital (AO) basis by running PySCF python package. }
    \label{tab:qub_red}
\end{table}

\section{Picking the right eigenvalues for the symmetries.}

Once we are able to construct all the symmetries, S of the Hamiltonian, we need to pick the right eigensector of the symmetry. This can be done using the relation between the $\sigma_z$ qubit operator and the occupation number in Jordan-Wigner transform:
\begin{align}
\sigma_z^i = a_i^{\dagger}a_i-1    
\end{align}
An occupied and unoccupied fermionic mode corresponds to the eigenvalue -1 and +1, of the $\sigma_z$ operator, respectively. This implies that a symmetry operator, $S$ which is a Pauli-Z string on say, a set $Q$ of qubits, is related to the parity operator of fermionic modes stored in those $Q$ qubits. In our case, we started with atomic orbitals and then transformed them with a $V$ matrix so as to get $R$ into a diagonal form. Consequently, the occupation numbers of these transformed orbitals are stored in the qubits. And based on whether the orbital is occupied or unoccupied in a given symmetry sector, it is possible to figure out the correct eigensector corresponding to the symmetry.

For example, in case of $BeH_2$ molecule, there are 14 orbitals in total with six electrons. The symmetries corresponding to conservation of spin up electrons and spin down electrons are 
$S_1=\sigma_z^1\sigma_z^2\sigma_z^3\sigma_z^4\sigma_z^5\sigma_z^6\sigma_z^7$ and $S_2=\sigma_z^8\sigma_z^9\sigma_z^{10}\sigma_z^{11}\sigma_z^{12}\sigma_z^{13}\sigma_z^{14}$, 
respectively. Since, there are six electrons in total, we know that each spin sector will have three electrons from the Hartree-Fock state. Thus, the correct sector for each of the symmetries, $S_1$ and $S_2$ is the one with $-1$ eigenvalue.

\emph{Molecular orbitals:} Most of the quantum chemistry software recognizes symmetries from geometry and constructs the molecular orbitals for the Hartree-Fock procedure accordingly. This means that molecular orbitals are already symmetrized corresponding to different irreps and that the symmetry operators will be ${Z}_2$ symmetries. One way to construct ${Z}_2$ symmetries will be to start from $R$ operators as presented in Section \ref{Sec:Trans_PG_to_SP} and then get the Pauli-Z string corresponding to the symmetry operator. The $R$ operator will require the knowledge of the molecular orbitals in terms of atomic orbitals.

The other method to obtain the symmetries will be to just run the subroutine presented in~\cite{Bravyi2017} to find the ${Z}_2$ symmetries. This process is little bit more efficient as it is automated whereas, currently it is required to manually construct the $R$-matrices in order to use the formalism presented in the present research. It may be possible to automate this but we leave this for the future work. One benefit of starting from $R$ matrices is that, one gets an intuitive understanding of the symmetries. The physical intuition allows us to tweak the details in order to have more symmetries in the system. It also serves as a verification procedure for debugging the software code as we already know the symmetries that we are expecting.

\section{Conclusions}
We have presented a formalism to exploit spatial symmetries present in molecular systems to reduce the number of qubits required for quantum simulations. Compared to one of the results presented in~\cite{Bravyi2017}, we find that additional symmetries can be discovered, and more qubits can be spared in the mapping to a quantum computer. The results presented here are relevant for optimizing resources in the context of NISQ computing~\cite{Preskill2018}.
    
\section{Acknowlegements}
KS and JDW are funded by NSF awards DMR-1747426, 1820747. JDW also acknowledges support from the U.S Department of Energy (Award A053685) and from Department of Energy-Office of Science, Office of Advanced Scientific Computing Research, under the Quantum Computing Application Teams program (Award 1979657). AM acknowledges support from the IBM Research Frontiers Institute. We acknowledge useful discussions with Sergey Bravyi.

\bibliography{bibfile}

\end{document}